\documentclass[runningheads]{llncs}
\usepackage[T1]{fontenc}
\usepackage{lmodern}
\usepackage{textcomp}
\usepackage{graphicx}
\usepackage{multirow}
\usepackage{amsfonts}
\usepackage{amsmath}
\usepackage{amssymb}
\usepackage{bookmark}
\usepackage{float}
\usepackage{listings}
\usepackage{graphicx}
\usepackage{cite}
\usepackage{subcaption}
\usepackage[ruled,linesnumbered]{algorithm2e}
\usepackage{booktabs}
\usepackage{hyperref}
\usepackage{xcolor}

\usepackage{xurl}
\usepackage{tikz}
\usetikzlibrary{shapes,arrows,shadows}

\newcommand{\orcid}[1]{\href{https://orcid.org/#1}{\includegraphics[width=10pt]{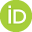}}}

\newcommand{\stateVal}{x}

\newcommand{\initialCondLabel}{\textit{Init}}
\newcommand{\nextCondLabel}{\textit{Next}}
\newcommand{\nextNncCondLabel}{\textit{Next}_\textit{NNC}}
\newcommand{\nextEnvCondLabel}{\textit{Next}_\textit{ENV}}

\newcommand{\nnFunction}{f_\textit{NN}}

\newcommand{\safeCondLabel}{\textit{Safe}}

\newcommand{\indInvLabel}{\textit{IndInv}}

\newcommand{\bridgePredicate}{\textit{Bridge}}

\newcommand{\reachableSet}[1]{\textit{Reach}(#1)}
\newcommand{\falsifyingPredicate}{\textit{FPred}}
\newcommand{\falsifyingState}{\textit{FState}}

\newcommand{\postConditionSet}[1]{\text{AllPosts}(#1)}
\newcommand{\nnPostCondition}{\psi}

\begin{document}

    \title{Compositional Inductive Invariant Based Verification of Neural Network Controlled Systems}
    \titlerunning{Compositional Inductive Invariant Based Verification of NNCSs}
    \author{Yuhao Zhou \orcid{0009-0003-6895-6308} \and 
    Stavros Tripakis \orcid{0000-0002-1777-493X}}
    \authorrunning{Y. Zhou and S. Tripakis}
    \institute{Northeastern University, Boston, MA, USA\\
    \email{\{zhou.yuhao, stavros\}@northeastern.edu}
    }

    \maketitle            

    \begin{abstract}
        The integration of neural networks into safety-critical systems has shown great potential in recent years. However, the challenge of effectively verifying the safety of Neural Network Controlled Systems (NNCS) persists. This paper introduces a novel approach to NNCS safety verification, leveraging the inductive invariant method. Verifying the inductiveness of a candidate inductive invariant in the context of NNCS is hard because of the scale and nonlinearity of neural networks. Our compositional method makes this verification process manageable by decomposing the inductiveness proof obligation into smaller, more tractable subproblems. Alongside the high-level method, we present an algorithm capable of automatically verifying the inductiveness of given candidates by automatically inferring the necessary decomposition predicates. The algorithm significantly outperforms the baseline method and shows remarkable reductions in execution time in our case studies, shortening the verification time from hours (or timeout) to seconds.

    \end{abstract}


    \section{Introduction}
    \label{sec:introduction}
    
{\em Neural Network Controlled Systems} (NNCSs) are closed-loop systems that consist of an {\em environment} controlled by a {\em neural network} (NN).
Advancements in machine learning have made NNCSs more widespread in safety-critical applications, such as autonomous cars, industrial control, and healthcare~\cite{zhang2020testingSurvey}. 
While many methods exist for the formal verification of standard closed-loop systems~\cite{clarke2018handbook}, the scale and nonlinearity of NNs makes the direct application of these methods to NNCSs a challenging task. 

In this paper, we study formal verification of safety properties of NNCSs, using the {\em inductive invariant method}~\cite{MannaPnueli95}. In a nutshell, the method consists in coming up with a state predicate $\indInvLabel$ which is {\em inductive}, an {\em invariant}, and implies safety (see Section~\ref{sec:prelim_problem_statement}, 
Conditions~(\ref{eq:indinv_initial}),~(\ref{eq:indinv_inductiveness}) and~(\ref{eq:indinv_safety})).
The most important property, and often the most difficult to check, is inductiveness. In the case of NNCSs, checking inductiveness amounts to checking the validity of the following formula,
where $\nextNncCondLabel$ and $\nextEnvCondLabel$ are the transition relation predicates of the NN controller and of the environment, respectively, and $\phi'$ is the state predicate $\phi$ applied to the next-state (primed) variables:
\begin{equation}
    (\indInvLabel \land \nextNncCondLabel \land \nextEnvCondLabel) \implies \indInvLabel'
  \label{eqinductivenessNNCS}
\end{equation}
The problem that motivates our work is that checking formula~(\ref{eqinductivenessNNCS})
is infeasible in practice using state-of-the-art tools. On the one hand, generic tools such as SMT solvers~\cite{z3} cannot handle formula~\ref{eqinductivenessNNCS} directly, typically due to the size and complexity of the NN and corresponding $\nextNncCondLabel$ predicate. On the other hand, specialized NN verification tools~\cite{wang2021beta,xu2020autolirpa} also cannot handle formula~(\ref{eqinductivenessNNCS})
directly, typically due to limitations in being able to deal with closed-loop systems and the environment transition relation $\nextEnvCondLabel$ (which, for example, might be non-deterministic).

This paper proposes a {\em compositional} method to deal with this problem. The key idea is the following: instead of checking 
formula~(\ref{eqinductivenessNNCS}) 
{\em monolithically}, we propose to break it up into two separate conditions:
\begin{align}
    & (\indInvLabel \land \nextNncCondLabel) \implies \bridgePredicate \quad  \label{eqnnc_check}\\
    & (\bridgePredicate \land \nextEnvCondLabel) \implies \indInvLabel'\quad \label{eqenv_check}
\end{align}
where $\bridgePredicate$ is a new state predicate that needs to be invented (we propose a technique to do this automatically).
Not only are each of the formulas~(\ref{eqnnc_check}) and~(\ref{eqenv_check})
smaller than the monolithic formula~(\ref{eqinductivenessNNCS}) 
they can also be handled by the corresponding specialized tool: 
formula~(\ref{eqnnc_check}) by a NN verifier, and 
formula~(\ref{eqenv_check}) by an SMT solver.

In Section~\ref{sec:our_approach} we elaborate our approach. We propose a technique to construct $\bridgePredicate$ predicates automatically, and show how this technique can also be used to establish the validity of 
Condition~(\ref{eqnnc_check})
by construction. We also present a heuristic that can falsify inductiveness compositionally in some cases when 
formula~(\ref{eqinductivenessNNCS})
is not valid. We incorporate these techniques into an algorithm that checks inductiveness of candidate $\indInvLabel$ predicates for NNCSs.

Our experimental results (Section~\ref{sec:experiments_implementation})
indicate that our algorithm consistently terminates, effectively verifying or falsifying the inductiveness of the given candidate in both deterministic and non-deterministic environment setups, for NNs of sizes up to $2\times 1024$ neurons, in a matter of seconds. This is in contrast to the monolithic method, which times out after one hour for all our experiments with NNs larger than $2\times 56$ neurons.

    \section{Preliminaries and Problem Statement}
    \label{sec:prelim_problem_statement}
    \subsubsection{Symbolic Transition Systems:}
\label{subsubsec:symbolic_transition_system}
The systems considered in this paper can be modeled as {\em symbolic transition systems} (STSs), like the ones shown in Figure~\ref{fig:sts-nnsts-example}. An STS is defined by: (1) a set of \textit{state variables}, each of appropriate type; (2) a  predicate $\initialCondLabel$ over the state variables, specifying the set of possible {\em initial states}; and (3) a predicate $\nextCondLabel$ over current and next (primed) state variables, specifying the {\em transition relation} of the system. 
For example, the STS shown in Figure~\ref{fig:mod-4-sts-example} models a counter modulo-4.
This system has one state variable $s$ of type integer ($\mathbb{Z}$). 
Its $\initialCondLabel$ predicate specifies that $s$ is initially $0$. 
$\nextCondLabel$ specifies that at each transition, $s$ is incremented by one until we reach $s=3$, upon which $s$ is reset to $0$ (primed variable $s'$ denotes the value of state variable $s$ at the next state).
The $\safeCondLabel$ predicate defines the set of {\em safe} states (used later).

A \textit{state} is an assignment of values to all state variables. 
We use the notation $\stateVal \models P$ to denote that state $\stateVal$ satisfies state
predicate $P$, i.e., that $P$ evaluates to true once we replace all state variables in $P$ by their values as given by $\stateVal$.
A {\em transition} is a pair of states $(\stateVal, \stateVal')$ that satisfies the predicate \nextCondLabel, denoted $(\stateVal, \stateVal')\models\nextCondLabel$. We also use notation $\stateVal \rightarrow \stateVal'$ instead of $(\stateVal, \stateVal')\models\nextCondLabel$, when $\nextCondLabel$ is clear from context.
A \textit{trace} is an infinite sequence of states $\stateVal_0, \stateVal_1, \cdots$ such
that $\stateVal_0 \models \initialCondLabel$ and $\stateVal_i \rightarrow \stateVal_{i+1}$ for all $i \ge 0$. A state $\stateVal$ is \textit{reachable} if there exists a trace $\stateVal_0, \stateVal_1, \cdots$, such that $\stateVal = \stateVal_i$ for some $i$. We use $\reachableSet{M}$ to denote the set of all reachable states of an STS $M$.

\begin{figure}[t]
  \centering
  \begin{subfigure}{0.35\textwidth}
    \centering
    \begin{align*}
      &\mbox{State variable}:  s \in \mathbb{Z} \\
      &\initialCondLabel: s = 0 \\
      &\nextCondLabel: (s = 3 \implies s' = 0) \;\; \land \\
      &  \qquad \;\; (s \neq 3 \implies s' = s + 1)\\
      &\safeCondLabel: 0 \le s \le 3
    \end{align*}
    \caption{Mod-4 counter}
    \label{fig:mod-4-sts-example}
  \end{subfigure}
  \hfill
  \begin{subfigure}{0.5\textwidth}
      \centering
      \begin{align*}
        &\mbox{State variable}:  s \in \mathbb{Z} \\
        &\initialCondLabel:  s = 0 \\
        &\mbox{NN function}: \nnFunction : \mathbb{Z} \rightarrow \mathbb{R} \\
        &\nextNncCondLabel: a = \nnFunction(s) \\
        &\nextEnvCondLabel: (a \ge 0 \implies s' = 0) \;\; \land \\
        &  \qquad \qquad \;\; (a < 0 \implies s' = s + 1)\\
        &\nextCondLabel: \nextNncCondLabel \land \nextEnvCondLabel \\
        &\safeCondLabel: 0 \le s \le 3
      \end{align*}
      \caption{Mod-4 counter with NN controller}
      \label{fig:mod-4-nnsts-example}
  \end{subfigure}
  \caption{Two STSs. The one in Fig.~\ref{fig:mod-4-nnsts-example} is a NNCS: function $\nnFunction$ (defined elsewhere) models a neural network controller.}
  \label{fig:sts-nnsts-example}
\end{figure}

\subsubsection{Invariants:}
\label{subsubsec:invariant}
In this paper we are interested in the verification of safety properties, and in particular \textit{invariants}, which
are state predicates that hold at all reachable states.
Formally, for an STS $M$, an invariant is a state predicate $\textit{Inv}$ such that $\reachableSet{M} \subseteq \{s \mid s \models \textit{Inv}\}$.
Predicates $s \ge 0$ and $0 \le s \le 3$ are both invariants of the STS of Figure~\ref{fig:mod-4-sts-example}. 

\subsubsection{Inductive Invariants:}
\label{subsubsec:inductive_invariant}

A standard technique for checking whether a given state predicate $\safeCondLabel$ is an invariant of a given STS is to come up with an {\em inductive invariant} stronger than $\safeCondLabel$, that is, with  
a state predicate $\indInvLabel$ which satisfies the following conditions~\cite{MannaPnueli95}:
\begin{align}
  \initialCondLabel \implies \indInvLabel \label{eq:indinv_initial} \\
  (\indInvLabel \land \nextCondLabel) \implies \indInvLabel' \label{eq:indinv_inductiveness}\\
  \indInvLabel \implies \safeCondLabel \label{eq:indinv_safety}
\end{align}
where $\indInvLabel'$ denotes the predicate $\indInvLabel$ where state variables are replaced by their primed, next-state versions.
Condition~(\ref{eq:indinv_initial}) states that $\indInvLabel$ holds at all initial states. Condition~(\ref{eq:indinv_inductiveness}) states that $\indInvLabel$ is \textit{inductive}, that is, if $\indInvLabel$ holds at a state $s$, then it also holds at any successor of $s$. We call Condition~(\ref{eq:indinv_inductiveness}) the \textit{inductiveness condition}. Condition~(\ref{eq:indinv_safety}) states that $\indInvLabel$ is stronger than $\safeCondLabel$. 
Conditions~(\ref{eq:indinv_initial}) and~(\ref{eq:indinv_inductiveness}) imply that all reachable states satisfy $\indInvLabel$, which together with Condition~(\ref{eq:indinv_safety}) implies that they also satisfy $\safeCondLabel$.

\subsubsection{Neural Networks:}
\label{subsubsec:neural_network}

At its core, a {\em neural network} (NN) is a function, receiving inputs and producing outputs. 
Mathematically, the output $z^{(i)}$ of each layer in a feed-forward NN with $L$ layers can be expressed as:
\begin{align*}
z^{(i)} = \sigma_i(W^{(i)}z^{(i-1)} + b^{(i)}), \quad i = 1, 2, \cdots, L,
\end{align*}
where $W^{(i)}$ and $b^{(i)}$ represent the weight matrix and bias vector for the $i$-th layer, respectively, and $\sigma_i$ is the activation function for that layer. Our paper considers feed-forward NNs with ReLU activation functions, where $\text{ReLU}(x) = \max(0, x)$.

\subsubsection{Neural Network Controlled Systems:}
\label{subsubsec:closed_loop_system}

In this paper, we are interested in safety verification for {\em neural network controlled systems} (NNCSs).
A NNCS is a STS consisting of a {\em neural network controller} and an {\em environment}, in closed-loop configuration.
Formally, a NNCS is a STS whose transition relation predicate $\nextCondLabel$ is of the form
$\nextNncCondLabel \land \nextEnvCondLabel$, where $\nextNncCondLabel$ is a predicate capturing the NN controller, and 
$\nextEnvCondLabel$ is a predicate capturing the transition relation  of the environment.
We assume that $\nextNncCondLabel$ is always of the form $\vec{a} = \nnFunction(\vec{s})$, where $\nnFunction$ is the function modeling the NN controller, $\vec{s}$ is the vector of state variables of the environment (which are the inputs to the NN controller), and $\vec{a}$ is the vector of outputs of the NN controller (which are the inputs to the environment). 
Then, $\nextEnvCondLabel$ is a predicate on $\vec{s}$, $\vec{s'}$, and $\vec{a}$.
Note that the only state variables in the NNCS are $\vec{s}$. Variables $\vec{a}$ are not state variables, but just temporary variables that can be eliminated and replaced by $\nnFunction(\vec{s})$, according to the equation $\vec{a} = \nnFunction(\vec{s})$.
We assume that for any assignment of $\vec{s}$ and $\vec{a}$, there always exists an assignment of $\vec{s'}$ such that $\nextEnvCondLabel$ is satisfied; that is, the system is deadlock-free.

An example NNCS is shown in Figure~\ref{fig:mod-4-nnsts-example}. 
In this simple example, the output $a$ of the NN controller controls the environment to either reset the state variable $s$ to $0$ (if $a \ge 0$) or increment it by $1$ (if $a<0$).

\subsubsection{Safety Verification Problem for NNCS:}
\label{subsubsec:problem_statement_nncs}

The problem we study in this paper is the safety verification problem for NNCS, namely:
{\em given a NNCS $M$ and a safety predicate $\safeCondLabel$, check whether $\safeCondLabel$ is an invariant of $M$}.

    \section{Our Approach}
    \label{sec:our_approach}
    To solve the safety verification problem for NNCSs, we will use the inductive invariant method described in Section~\ref{sec:prelim_problem_statement}.
In particular, our approach assumes that a {\em candidate} inductive invariant is given, and our focus is to {\em check} whether this candidate is indeed a valid inductive invariant stronger than $\safeCondLabel$, i.e., whether it satisfies 
Conditions~(\ref{eq:indinv_initial}),~(\ref{eq:indinv_inductiveness}), and~(\ref{eq:indinv_safety}). Specifically, we focus on checking {\em inductiveness}, i.e., Condition~(\ref{eq:indinv_inductiveness}), because this is the most challenging condition to check.

In the case of NNCS, 
Condition~(\ref{eq:indinv_inductiveness}) instantiates to:
\begin{align}
    \label{eq:monolithic_method_inductiveness}
    (\indInvLabel \land \nextNncCondLabel \land \nextEnvCondLabel) \implies \indInvLabel'.
\end{align}
A naive approach is to attempt to check Condition~(\ref{eq:monolithic_method_inductiveness}) directly:
we call this the \textit{monolithic method}. 
Unfortunately, as we will show in Section~\ref{sec:experiments_implementation}, the monolithic method does not scale.
This is typically because of the size of the $\nextNncCondLabel$ part of the formula, which encodes the NN controller, and tends to be very large.
To address this challenge, we introduce a \textit{compositional method} for checking inductiveness, described next.

\subsection{Our Approach: Compositional Method}

Our compositional method is centered around two key ideas: 
(i) automatically construct a \textit{bridge predicate}, denoted $\bridgePredicate$; and (ii) replace the monolithic inductiveness Condition~(\ref{eq:monolithic_method_inductiveness}) by two separate conditions:
\begin{align}
    & (\indInvLabel \land \nextNncCondLabel) \implies \bridgePredicate \quad  \label{eq:nnc_check}\\
    & (\bridgePredicate \land \nextEnvCondLabel) \implies \indInvLabel'\quad \label{eq:env_check}
\end{align}
By transitivity of logical implication, it is easy to show the following:
\begin{theorem}[Soundness]
\label{thm:sufficiency_of_bridge_predicate}
If Conditions~(\ref{eq:nnc_check}) and~(\ref{eq:env_check}) hold then Condition~(\ref{eq:monolithic_method_inductiveness}) holds.
\end{theorem}

The completeness of our method follows from the fact that in the worst case 
we can set $\bridgePredicate$ to be equal to $\indInvLabel\land\nextNncCondLabel$.
\begin{theorem}[Completeness]
\label{thm:existence_of_bridge_predicate}
If Condition~(\ref{eq:monolithic_method_inductiveness}) holds then we can find $\bridgePredicate$ such that Conditions~(\ref{eq:nnc_check}) and~(\ref{eq:env_check}) hold.
\end{theorem}

As it turns out (c.f. Section~\ref{sec:experiments_implementation}) checking Conditions~(\ref{eq:nnc_check}) and~(\ref{eq:env_check}) separately scales much better than checking Condition~(\ref{eq:monolithic_method_inductiveness}) monolithically.
However, this relies on finding a ``good'' bridge precidate.
Setting $\bridgePredicate$ to $\indInvLabel\land\nextNncCondLabel$ is not helpful, because then Condition~(\ref{eq:env_check}) becomes identical to the monolithic Condition~(\ref{eq:monolithic_method_inductiveness}). 
Thus, a necessary step is finding a bridge predicate that satisfies Conditions~(\ref{eq:nnc_check}) and~(\ref{eq:env_check}), while remaining manageable. In Section~\ref{subsec:automatic_bridge_predicate_inference}, we present an automatic technique for doing so. But first, we examine bridge predicates in more depth.

\subsubsection{Naive Bridge Predicates are Incomplete:}
A natural starting point for constructing the bridge predicate is to define it only over the output variables $\vec{a}$ of the NN controller.
We call this a \textit{naive bridge predicate}.
Unfortunately, naive bridge predicates might be insufficient, as we show next.
Consider the system shown in Figure~\ref{fig:mod-4-nnsts-example}. Suppose we set $\nextNncCondLabel$ to: 
\begin{align*}
    ((0 \le s\le 2) \land a = -1) \lor (s = 3 \land a = 1) \lor ((s < 0 \lor s > 3 ) \land  (a=-1 \lor a=1))
\end{align*}
It can be checked that the set of reachable states of this system is $0 \le s \le 3$.
Therefore, $0 \le s \le 3$ is also an inductive invariant of this system.
A naive bridge predicate that satisfies Condition~(\ref{eq:nnc_check}) could be $a = 1 \lor a = -1$. 
This is, in fact, the strongest naive bridge predicate that satisfies Condition~(\ref{eq:nnc_check}), as it captures 
all the possible values of $a$.
Therefore, if this naive bridge predicate does not satisfy Condition~(\ref{eq:env_check}), then no naive bridge predicate that satisfies both Conditions~(\ref{eq:nnc_check}) and~(\ref{eq:env_check}) exists. Indeed, 
Condition~(\ref{eq:env_check}) is violated by this naive bridge, as it becomes
\begin{align*}
((a=1 \lor a=-1) \land (a \ge 0 \implies s' = 0) \land (a < 0 \implies s' = s + 1))\\
\quad \implies (0 \le s' \le 3)
\end{align*}
which is false when $s= 3$, $a = -1$ and $s'=4$. So, no naive bridge predicate can satisfy both Condition~(\ref{eq:nnc_check}) and~(\ref{eq:env_check}).
This suggests that naive bridge predicates are insufficient. 
The solution is to allow bridge predicates to refer both to the inputs and outputs of the NN controller:

\subsubsection{Generalized Bridge Predicates are Complete:}
A {\em generalized bridge predicate} is a predicate defined over both the output variables $\vec{a}$ as well as the input variables $\vec{s}$ of the NN controller (the inputs $\vec{s}$ are the same as the state variables of the environment). 
Continuing our example above, a generalized bridge predicate could be: $((0 \le s\le 2) \land a = -1) \lor (s = 3 \land a = 1)$.
Then, Conditions~(\ref{eq:nnc_check}) and~(\ref{eq:env_check}) become:

\begin{align}
 & \nonumber ((0 \le s \le 3) \quad \land \\
 & \nonumber ((0 \le s\le 2) \land a = -1) \lor (s = 3 \land a = 1) \lor ((s < 0 \lor s > 3 ) \land  (a=-1 \lor a=1)))\\
 & \implies (((0 \le s\le 2) \land a = -1) \lor (s = 3 \land a = 1))
\end{align}
\begin{align}
& \nonumber (((0 \le s\le 2) \land a = -1) \lor (s = 3 \land a = 1)) 
\land   (a \ge 0 \implies s' = 0)\\
& \nonumber \land ~ (a < 0 \implies s' = s + 1))\\
&\implies (0 \le s' \le 3).
\end{align}
and it can be checked that both are valid, which shows that this generalized bridge predicate is sufficient. 

In general, and according to Theorem~\ref{thm:existence_of_bridge_predicate}, a generalized bridge predicate is sufficient, since we can set $\bridgePredicate$ to $\indInvLabel\land\nextNncCondLabel$, and the latter predicate is over $\vec{s}$ and $\vec{a}$. But it is important to note that a bridge that works need not be the ``worst-case scenario'' predicate $\indInvLabel\land\nextNncCondLabel$. Indeed, the bridge of our example is not, and neither are the bridges constructed by our tool for the case studies in Section~\ref{sec:experiments_implementation}.

\subsection{Automatic Inference of Generalized Bridge Predicates}
\label{subsec:automatic_bridge_predicate_inference}

The key idea of our automatic bridge inference technique is to synthesize a generalized bridge predicate 
such that Condition~(\ref{eq:nnc_check}) holds {\em by construction} (i.e., it does not need to be checked).
To explain how this works, let us first define, given a predicate $A$ over both $\vec{s}$ and $\vec{a}$, and predicate $B$ over only $\vec{s}$, the {\em set of postconditions}, denoted 
$\postConditionSet{A, B}$, to be the set of all predicates $C$ over $\vec{a}$ such that $(A \land B) \implies C$ is valid. 

Then, our algorithm will construct a bridge predicate the has the  form:
\begin{align}
    \bridgePredicate = 
	(p_1 \land \nnPostCondition_1) \lor (p_2 \land \nnPostCondition_2) \lor \cdots \lor (p_n \land \nnPostCondition_n)
	\label{eq:bridge_predicate_form}
\end{align}
where each $p_i$ is a predicate over $\vec{s}$, and each $\psi_i$ is a predicate over $\vec{a}$, such that the following conditions hold: 
\begin{align}
    \label{eq:ind_inv_split_condition}
    & (p_1\lor p_2 \lor \cdots \lor p_n) \iff \indInvLabel\\
    \label{eq:nnc_post_condition_in_bridge}
    & \forall i=1,...,n, \;\; \nnPostCondition_i \in \postConditionSet{\nextNncCondLabel, p_i}
\end{align}
Condition~(\ref{eq:ind_inv_split_condition}) ensures that the set of all $p_i$'s is a {\em decomposition} of the candidate inductive invariant $\indInvLabel$, i.e., that their union, viewed as sets, ``covers'' $\indInvLabel$.
This decomposition need not be a partition of $\indInvLabel$, i.e., the $p_i$'s need not be disjoint.
Condition~(\ref{eq:nnc_post_condition_in_bridge}) ensures that each $\nnPostCondition_i$ is a postcondition of the corresponding $p_i$ w.r.t. the NN controller (note that $\nnPostCondition_i$ is not necessarily the strongest postcondition).

Now, from~(\ref{eq:bridge_predicate_form}) and~(\ref{eq:ind_inv_split_condition}), Condition~(\ref{eq:nnc_check}) becomes:
\begin{align*}
    ((p_1\lor p_2 \lor \cdots\lor p_n) \land \nextNncCondLabel) \implies
    ((p_1 \land \nnPostCondition_1) \lor (p_2 \land \nnPostCondition_2) \lor \cdots\lor (p_n \land \nnPostCondition_n))
\end{align*}
which is equivalent to
\begin{align}
    \left(\bigvee_{i=1}^n (p_i \land \nextNncCondLabel)\right) \implies \left(\bigvee_{j=1}^n (p_j \land \nnPostCondition_j)\right) 
\end{align}
which is equivalent to
\begin{align}
	\label{condconstruction}
    \bigwedge_{i=1}^n \left((p_i \land \nextNncCondLabel) \implies (\bigvee_{j=1}^n (p_j \land \nnPostCondition_j))\right)
\end{align}
But observe that $(p_i \land \nextNncCondLabel) \implies p_i$ trivially holds, for each $i=1,...,n$.
And note that $(p_i \land \nextNncCondLabel) \implies \nnPostCondition_i$ also holds, since by construction,
$\nnPostCondition_i \in \postConditionSet{\nextNncCondLabel, p_i}$. Therefore,
$(p_i \land \nextNncCondLabel) \implies (p_i \land \nnPostCondition_i)$ holds for each $i$,
which implies that~(\ref{condconstruction}) holds by construction.
Therefore, we have:

\begin{theorem}[Condition~(\ref{eq:nnc_check}) holds by construction]
    \label{thm:nnc_check_valid_by_construction}
    For any bridge predicate that is in form of Condition~(\ref{eq:bridge_predicate_form}), if this predicate satisfies Conditions~(\ref{eq:ind_inv_split_condition}) and~(\ref{eq:nnc_post_condition_in_bridge}), then Condition~(\ref{eq:nnc_check}) holds by construction.
\end{theorem}

Next, consider Condition~(\ref{eq:env_check}).
From~(\ref{eq:bridge_predicate_form}),  Condition~(\ref{eq:env_check}) becomes:

\begin{align}
    \Big(\big((p_1 \land \nnPostCondition_1) \lor (p_2 \land \nnPostCondition_2) \lor \cdots\lor (p_n \land \nnPostCondition_n)\big) \land \nextEnvCondLabel \Big) \implies
    \indInvLabel'
\end{align}
which is equivalent to:
\begin{align}
    \left(\bigvee_{i=1}^n (p_i \land \nnPostCondition_i \land \nextEnvCondLabel) \right) \implies \indInvLabel'
\end{align}
which is equivalent to:
\begin{align}
    \label{eq:env_check_clauses}
    \bigwedge_{i=1}^n \left((p_i \land \nnPostCondition_i \land \nextEnvCondLabel) \implies \indInvLabel'\right) 
\end{align}
which means that checking Condition~(\ref{eq:env_check}) can be replaced by checking $n$ smaller conditions, namely,
$(p_i \land \nnPostCondition_i \land \nextEnvCondLabel) \implies \indInvLabel'$, for $i=1,...,n$.

The algorithm that we present below (Algorithm~\ref{alg:inductiveness_check}) starts with $n=1$ and $p_1 = \indInvLabel$.
It then computes some $\nnPostCondition_1 \in \postConditionSet{\nextNncCondLabel, p_1}$ and checks whether
$(p_1 \land \nnPostCondition_1 \land \nextEnvCondLabel) \implies \indInvLabel'$ is valid.
If it is, then $p_1 \land \nnPostCondition_1$ is a valid bridge and inductiveness holds.
Otherwise, $p_1$ is {\em split} into several $p_i$'s, and the process repeats.

\subsection{Heuristic for Falsifying Inductiveness}
\label{sec:falsifying}

An additional feature of our algorithm is that it is often capable to {\em falsify} inductiveness and thereby prove that it does not hold. This allows us to avoid searching hopelessly for a bridge predicate when none exists, because the candidate invariant is not inductive. 
 
Directly falsifying Condition~(\ref{eq:monolithic_method_inductiveness})  faces the same scalability issues as trying to prove this condition monolithically. On the other hand, falsifying Conditions~(\ref{eq:nnc_check}) or~(\ref{eq:env_check}) is not sufficient to disprove inductiveness. The failure to prove these conditions might simply mean that our chosen bridge predicate does not work.

Therefore, we propose a practical heuristic which inherits the decomposition ideas described above. This heuristic involves constructing: 
(i) a satisfiable \textit{falsifying state predicate} over $\vec{s}$, denoted $\falsifyingState$; and (ii) a predicate over the output of the NN controller $\vec{a}$, denoted $\falsifyingPredicate$, 
such that the following conditions hold:
\begin{align}
\falsifyingState \implies \indInvLabel \label{eq:fal_state} \\ 
\label{eq:falsifying_nnc_check}
(\falsifyingState \land \nextNncCondLabel) \implies \falsifyingPredicate \\
\label{eq:falsifying_env_check}
(\falsifyingState \land \falsifyingPredicate \land \nextEnvCondLabel) \implies \neg \indInvLabel'
\end{align}
Intuitively, $\falsifyingState$ identifies a set of states which satisfy $\indInvLabel$ but violate $\indInvLabel'$ after a transition. $\falsifyingPredicate$ captures the outputs of the NN controller when its inputs satisfy $\falsifyingState$. The conjunction of~(\ref{eq:falsifying_nnc_check}) and~(\ref{eq:falsifying_env_check}) implies:
\begin{align}
\label{condfalsify}
(\falsifyingState \land \nextNncCondLabel \land \nextEnvCondLabel) \implies \neg \indInvLabel'
\end{align}
In turn, (\ref{condfalsify}) and~(\ref{eq:fal_state}) together imply that Condition~(\ref{eq:monolithic_method_inductiveness}) does not hold. This leads us to the following theorem:

\begin{theorem}[Falsifying Inductiveness]
    \label{thm:falsifying_predicate}
    If $\falsifyingState$ is satisfiable, and if Conditions~(\ref{eq:fal_state}), (\ref{eq:falsifying_nnc_check}), and (\ref{eq:falsifying_env_check}) hold, then Condition~(\ref{eq:monolithic_method_inductiveness}) does not hold.
\end{theorem}

Our falsification approach aligns well with the composite structure of a NNCS, and is compositional, because it allows to check separately the NN transition relation $\nextNncCondLabel$ in~(\ref{eq:falsifying_nnc_check}) and the environment transition relation $\nextEnvCondLabel$ in~(\ref{eq:falsifying_env_check}).

\subsection{Algorithm}
\label{subsec:algorithm}

The aforementioned ideas are combined in Algorithm~\ref{alg:inductiveness_check}, which implements our compositional inductiveness verification approach for NNCSs.
Line 7 ensures that $\nnPostCondition$ is a postcondition of $p$ w.r.t. the NN controller. 
In practice we use a {\em NN verifier} to compute the postcondition
(see Section~\ref{sec:experiments_implementation}). 
Lines 15 and 16 ensure Condition~(\ref{eq:ind_inv_split_condition}), together with the fact that the initial $p$ is $\indInvLabel$ (Line 3).
Line~9 ensures that the bridge predicate is in form of Condition~(\ref{eq:bridge_predicate_form}). Therefore, 
by Theorem~\ref{thm:nnc_check_valid_by_construction}, Algorithm~\ref{alg:inductiveness_check} ensures Condition~(\ref{eq:nnc_check})  by construction.

Line 8 corresponds to checking~(\ref{eq:env_check_clauses}) compositionally, i.e., separately for each $i$. In practice, we use an {\em SMT solver} for this check (see Section~\ref{sec:experiments_implementation}). 

For falsification, to use Theorem~\ref{thm:falsifying_predicate}, we 
set $\falsifyingState = p$ and $\falsifyingPredicate = \nnPostCondition$. 
The splitting process at Line~15 guarantees that $p$ is satisfiable.
Condition~(\ref{eq:fal_state}) then becomes $p \implies \indInvLabel$, which holds by construction because of Condition~(\ref{eq:ind_inv_split_condition}). Condition~(\ref{eq:falsifying_nnc_check}) becomes $p \land \nextNncCondLabel \implies \nnPostCondition$, which holds because $\nnPostCondition\in\postConditionSet{\nextNncCondLabel,p}$. 
Condition~(\ref{eq:falsifying_env_check}) becomes
    $(p \land \nnPostCondition \land \nextEnvCondLabel) \implies \neg \indInvLabel'$, 
which is checked in Line~11. 

If the algorithm can neither prove  $(p \land \nnPostCondition \land \nextEnvCondLabel) \implies \indInvLabel'$, nor falsify  inductiveness, then it splits $p$ into a disjunction of satisfiable state predicates, ensuring Condition~(\ref{eq:bridge_predicate_form}). Our implementation utilizes various splitting strategies. The specific splitting strategies employed in our case studies are elaborated in Section~\ref{sec:experiments_implementation}.
After splitting, all resulting state predicates are added into the queue $Q$. Consequently, the queue becomes empty if and only if the validity of every conjunct in Condition~(\ref{eq:env_check_clauses}) is proved. Therefore, an empty queue indicates that  inductiveness holds.

\begin{algorithm}[t]
\caption{Compositional Inductiveness Checking for NNCS}
\label{alg:inductiveness_check}
\KwIn{Transition relations $\nextNncCondLabel$, $\nextEnvCondLabel$; Candidate Inductive Invariant $\indInvLabel$}
\KwOut{Verification result: (True with bridge predicate $Bridge$) or (False with falsifying state predicate)}
\SetKwFunction{FMain}{CheckInductiveness}
  \SetKwProg{Fn}{Function}{:}{}
  \Fn{\FMain{$\nextNncCondLabel$, $\nextEnvCondLabel$, $\indInvLabel$}}{
    $\bridgePredicate := False$ \; 
    $Q := \{\indInvLabel\}$ \;
    \While{$Q \ne \emptyset$}{
        Let $p \in Q$ \;
        $Q := Q \setminus \{p\}$ \;
        Let $\nnPostCondition \in \postConditionSet{\nextNncCondLabel, p}$ \;
        \If{$(p \land \nnPostCondition \land \nextEnvCondLabel) \implies \indInvLabel'$ holds}{
            $\bridgePredicate := \bridgePredicate \lor (p \land \nnPostCondition)$ \;
        }
        \ElseIf{$(p \land \nnPostCondition \land \nextEnvCondLabel) \implies \neg \indInvLabel'$ holds}{
            \KwRet{(False, $p$)} \;
        }
        \Else{
            Split $p$ into $p_1, p_2, \cdots, p_k$ such that  $p\iff (p_1\lor p_2 \lor \cdots\lor p_k)$;\\
            $Q := Q \cup \{p_1, p_2, \cdots , p_k\}$ \;
        }
    }
    \KwRet{(True, $\bridgePredicate$)} \;
  }
\end{algorithm}

\noindent
{\bf Termination:}
the algorithm terminates either upon successfully proving or upon falsifying inductiveness. However, termination is not guaranteed in all cases. 
In infinite state spaces, the algorithm may keep splitting predicates ad infinitum.
In practice,  as observed in our case studies, the algorithm consistently terminated (see Section~\ref{sec:experiments_implementation}).

    \section{Evaluation}
    \label{sec:experiments_implementation}

In the experiments reported below, we evaluate our compositional method by comparing it against the monolithic method
which uses Z3 to check Condition~\ref{eq:monolithic_method_inductiveness} directly.
We remark that we also attempted to use 
specialized NN verifiers such as~\cite{wang2021beta}
to check inductiveness monolithically, but this proved infeasible.
Specifically, we tried two ways
of encoding checking inductiveness as an NN verification problem: 
(i) Encoding $\nextEnvCondLabel$ and Condition~(\ref{eq:monolithic_method_inductiveness}) into the NN's input and output constraints.  
This approach was impractical for our case studies since $\nextEnvCondLabel$ involves arithmetic applied simultaneously to both the input and output of the NN controller, e.g., $(x' = x + 0.1 a)$. Such constraints are beyond the capability of existing NN verifiers such as~\cite{katzReluplexEfficientSMT2017,katz2019marabou,wang2021beta,bak2021nfm}. 
(ii) Encoding both the NN controller and the environment into a single NN. 
Our trials revealed that current NN verifiers support only a limited range of operators for defining a NN, restricting this approach as well. Notably, the operators we required, such as adding the NN's input to its output, are not typically supported. Although these limitations might diminish as NN verifiers evolve, an additional complication arises from the fact that $\nextEnvCondLabel$ might be non-deterministic (as in our second set of case studies). Encoding non-deterministic transition relations into a NN remains a significant challenge because NNs are typically deterministic functions. 

In our evaluation, we do not attempt to systematically compare our tool against NNCS reachability analysis tools, e.g.~\cite{fanReachNNToolReachability2020,verisig20,tran2020nnv,SchillingFG22julia_reach_nncs}, primarily because these tools perform verification over a bounded time horizon through reachability analysis, while our method is based on inductive invariants and enables the verification of safety properties over an infinite time horizon.
However, 
following the recommendation of an anonymous Reviewer,
we did execute some of the aforementioned tools on the 2D maze case studies described in Section~\ref{sec2Dmazes} that follows. 
In summary, in many of our experiments, tools such as JuliaReach~\cite{bogomolov2019juliareach} and NNV~\cite{tran2020nnv} have been able to successfully verify safety within a bounded time horizon. 
For example, it took NNV a few seconds to perform a 50-step reachability analysis on the deterministic 2D maze with the largest NN controller that we used ($2\times 1024$ neurons).
However, both JuliaReach and NNV also failed to verify safety in some instances, due to the overapproximation of the reach set. 
Our experiments with JuliaReach and NNV are reported in Appendix~\ref{sec:appendix_other_tools_experiments}.

\subsection{Implementation and Experimental Setup}

We implemented Algorithm~\ref{alg:inductiveness_check} in a prototype tool --
the source code and models needed to reproduce our experiments are available at~\url{https://github.com/YUH-Z/comp-indinv-verification-nncs}.
We use the SMT solver Z3~\cite{z3} for the validity checks of Lines~8 and~11 of Algorithm~\ref{alg:inductiveness_check}. 
To compute the postcondition $\nnPostCondition$ (Line~7 of Algorithm~\ref{alg:inductiveness_check}), we use AutoLIRPA~\cite{xu2020autolirpa}, 
which is the core engine of the NN verifier $\alpha$-$\beta$-CROWN~\cite{wang2021beta}.
AutoLIRPA computes a postcondition $\nnPostCondition\in\postConditionSet{\nextNncCondLabel, p}$, i.e., guarantees that 
$(p\land\nextNncCondLabel)\implies \nnPostCondition$, but it does not generally guarantee that $\nnPostCondition$ is the strongest possible postcondition.

In our case studies reported below, each candidate inductive invariant is a union of hyperrectangles, which aligns with the constraint types supported by mainstream NN verifiers such as~\cite{katz2019marabou,katzReluplexEfficientSMT2017,bak2021nfm,wang2021beta}.

Our splitting strategy (Line~15 of Algorithm~\ref{alg:inductiveness_check}) follows a binary scheme, dividing hyperrectangles at their midpoints along each dimension. For example, the interval $[1, 2]$ would be split into $[1, 3/2]$ and $[3/2, 2]$. This straightforward strategy turns out to be effective in our case studies.

Our experiments were conducted on a machine equipped with a 3.3 GHz 8-core AMD CPU, 16 GB memory, and an Nvidia 3060 GPU. Z3 ran on the CPU and AutoLIRPA on the GPU with CUDA enabled, both using default configurations.

\subsection{Case Studies and Experimental Results}
\label{sec2Dmazes}

We ran our experiments on two sets of examples, a {\em deterministic 2D maze}, and a {\em non-deterministic 2D maze}, as described below.

\textbf{Deterministic 2D maze:} 
In this example, the environment has two state variables $x$ and $y$ of type real ($\mathbb{R}$), representing an object's 2D position. 
The NN controller $\nnFunction : \mathbb{R}^2 \rightarrow \mathbb{R}^2$ outputs $(a, b)=\nnFunction(x, y)$, guiding the object's horizontal and vertical movement.
The initial state is set within $0.3 \le x \le 0.4$ and $0.6 \le y \le 0.7$.
The transition relation $\nextEnvCondLabel$ of the environment is deterministic and defined by $x' = x + 0.1a$ and $y' = y + 0.1b$.
The controller's goal is to navigate the object to ($0.8 \le x \le 0.9$ and $0.8 \le y \le 0.9$), while keeping it within a safe region of ($0.22 \le x \le 0.98$ and $0.54 \le y \le 0.98$), which is the safety property we want to prove.

As NN controllers, we used two-layer feed-forward NNs with ReLU activations, trained via PyTorch \cite{pytorch-ref} and Stable-Baselines3 \cite{stable-baselines3}. 
We trained NN controllers of various sizes, ranging from $2 \times 32$ to $2 \times 1024$ neurons. 
To standardize the comparison among the systems containing different NN controllers, for each system, we check the same candidate inductive invariant, defined as $0.25 \le x \le 0.95$ and $0.55 \le y \le 0.95$. For this candidate and for the $\initialCondLabel$ and $\safeCondLabel$ predicates defined above, it is easy to see that Conditions~(\ref{eq:indinv_initial}) and~(\ref{eq:indinv_safety}) hold. So our experiments only check the inductiveness Condition~(\ref{eq:monolithic_method_inductiveness}).

\begin{table*}[t]
  \centering
  \caption{Experimental results: {\em Det} and {\em NDet} indicate the results for the deterministic and non-deterministic 2D maze case studies, respectively. All execution times are in seconds. T.O. represents an one-hour timeout. The {\em Verified?} column shows whether the compositional method successfully terminated, either by proving (T) or by disproving (F) inductiveness.  {\em \#Splits} reports the total number of splits performed. {\em \#SMT} and {\em \#NNV queries} report the total number of calls made to Z3 and AutoLIRPA, respectively.
    }
  \label{tab:2d-maze-1}
\begin{tabular}{|l|ll|ll|ll|ll|ll|ll|}
\hline
\multirow{2}{*}{NN size} & \multicolumn{2}{l|}{Verified?}   & \multicolumn{2}{l|}{\begin{tabular}[c]{@{}l@{}}Monolithic \\execution time\end{tabular}} & \multicolumn{2}{l|}{\begin{tabular}[c]{@{}l@{}}Compositional \\execution time\end{tabular}} & \multicolumn{2}{l|}{\#Splits}   & \multicolumn{2}{l|}{\begin{tabular}[c]{@{}l@{}}\#SMT \\queries\end{tabular}} & \multicolumn{2}{l|}{\begin{tabular}[c]{@{}l@{}}\#NNV \\queries\end{tabular}} \\ \cline{2-13} 
                          & \multicolumn{1}{l|}{Det} & NDet & \multicolumn{1}{l|}{Det}                       & NDet                      & \multicolumn{1}{l|}{Det}                       & NDet                      & \multicolumn{1}{l|}{Det} & NDet & \multicolumn{1}{l|}{Det}                        & NDet                        & \multicolumn{1}{l|}{Det}                        & NDet                        \\ \hline
$2\times 32$              & \multicolumn{1}{l|}{T}   & T    & \multicolumn{1}{l|}{51.59}                     & 39.73                     & \multicolumn{1}{l|}{0.73}                      & 0.83                      & \multicolumn{1}{l|}{12}  & 14   & \multicolumn{1}{l|}{61}                         & 71                          & \multicolumn{1}{l|}{49}                         & 57                          \\ \hline
$2\times 40$              & \multicolumn{1}{l|}{T}   & T    & \multicolumn{1}{l|}{113.69}                    & 296.61                    & \multicolumn{1}{l|}{0.70}                      & 0.66                      & \multicolumn{1}{l|}{13}  & 13   & \multicolumn{1}{l|}{66}                         & 66                          & \multicolumn{1}{l|}{53}                         & 53                          \\ \hline
$2\times 48$              & \multicolumn{1}{l|}{T}   & T    & \multicolumn{1}{l|}{410.14}                    & 3002.20                   & \multicolumn{1}{l|}{0.52}                      & 0.42                      & \multicolumn{1}{l|}{10}  & 8    & \multicolumn{1}{l|}{51}                         & 41                          & \multicolumn{1}{l|}{41}                         & 33                          \\ \hline
$2\times 56$              & \multicolumn{1}{l|}{T}   & T    & \multicolumn{1}{l|}{1203.76}                   & T.O.                      & \multicolumn{1}{l|}{0.42}                      & 0.46                      & \multicolumn{1}{l|}{8}   & 9    & \multicolumn{1}{l|}{41}                         & 46                          & \multicolumn{1}{l|}{33}                         & 37                          \\ \hline
$2\times 64$              & \multicolumn{1}{l|}{T}   & T    & \multicolumn{1}{l|}{T.O.}                      & T.O.                      & \multicolumn{1}{l|}{0.76}                      & 0.61                      & \multicolumn{1}{l|}{15}  & 12   & \multicolumn{1}{l|}{76}                         & 61                          & \multicolumn{1}{l|}{61}                         & 49                          \\ \hline
$2\times 128$             & \multicolumn{1}{l|}{F}   & T    & \multicolumn{1}{l|}{T.O.}                      & T.O.                      & \multicolumn{1}{l|}{2.21}                      & 1.43                      & \multicolumn{1}{l|}{64}  & 28   & \multicolumn{1}{l|}{225}                        & 141                         & \multicolumn{1}{l|}{160}                        & 113                         \\ \hline
$2\times 256$             & \multicolumn{1}{l|}{T}   & T    & \multicolumn{1}{l|}{T.O.}                      & T.O.                      & \multicolumn{1}{l|}{1.68}                      & 1.18                      & \multicolumn{1}{l|}{27}  & 23   & \multicolumn{1}{l|}{136}                        & 116                         & \multicolumn{1}{l|}{109}                        & 93                          \\ \hline
$2\times 512$             & \multicolumn{1}{l|}{T}   & T    & \multicolumn{1}{l|}{T.O.}                      & T.O.                      & \multicolumn{1}{l|}{3.04}                      & 2.30                      & \multicolumn{1}{l|}{60}  & 45   & \multicolumn{1}{l|}{301}                        & 226                         & \multicolumn{1}{l|}{241}                        & 181                         \\ \hline
$2\times 1024$            & \multicolumn{1}{l|}{T}   & T    & \multicolumn{1}{l|}{T.O.}                      & T.O.                      & \multicolumn{1}{l|}{1.94}                      & 5.23                      & \multicolumn{1}{l|}{38}  & 102  & \multicolumn{1}{l|}{191}                        & 511                         & \multicolumn{1}{l|}{153}                        & 409                         \\ \hline
\end{tabular}
\end{table*}

The results are reported in Table~\ref{tab:2d-maze-1}: for this case study, the relevant columns are those marked as {\em Det}.
The compositional method successfully terminated in all cases, and proved inductiveness for all NN configurations, except for $2\times 128$, for which inductiveness does not hold, and in which case the compositional method managed to falsify it. 
In terms of performance, the monolithic method requires around one minute
even for the smallest NN configurations, and times out after one hour for the larger configurations; while the compositional method terminates in all cases in a couple of seconds. 
A key metric of the compositional method is the number of splits, which is determined by the number of times  Line~15 of Algorithm~\ref{alg:inductiveness_check} is executed.
This metric is crucial as it {impacts} the number of queries made to the SMT solver and to the NN verifier, typically the most time-consuming steps in the algorithm. Additionally, the number of splits indicates the size of the bridge predicate, where fewer splits suggest smaller bridge predicates.

\textbf{Non-deterministic 2D maze:} 
This case study is similar to the deterministic 2D maze, with the difference that the environment's transition relation is non-deterministic, specifically, defined as $x' = x + 0.1 c\cdot a$ and $y' = y + 0.1 c\cdot b$, where $c$ is a constant of type real, non-deterministically ranging between 0.5 and 1.0. 
This non-determinism simulates the noise within the system. 

The NN controllers in this set of experiments, while maintaining the same architecture and size as in the deterministic case studies, were retrained from scratch to adapt to the new transition relation. The candidate inductive invariant is the same as in the deterministic case study, ensuring consistency in our comparative analysis. 

The results are shown in Table~\ref{tab:2d-maze-1}, columns marked by {\em NDet}.
As demonstrated by the results,
our compositional method successfully verified 
inductiveness in all configurations, in a matter of seconds.

    \section{Related Work}
    \label{sec:related_work}
    A large body of research  exists on NN verification, including methods that verify NN input-output relations, using SMT solvers~\cite{huangSafetyVerificationDeep2017, katzReluplexEfficientSMT2017, katz2019marabou} or MILP solvers~\cite{Tjeng2019EvaluatingRO}, as well as methods that employ abstract-interpretation techniques~\cite{gehrAI2SafetyRobustness2018, singh2018fast}, symbolic interval propagation~\cite{wangshiqiReluVal2018}, dual optimization~\cite{dvijotham2018dual}, linear relaxation~\cite{xu2020autolirpa}, and bound propagation~\cite{wang2021beta}.
\cite{narodytska2018verifying_binarized_dnn} formally verifies equivalence and adversarial robustness of binarized NNs by SAT solvers.
\cite{NNequivFORMATS2022} propose an SMT based approach for checking equivalence and approximate equivalence of NNs.

These and other techniques and the corresponding tools focus on NN verification at the {\em component level}, that is, they verify a NN in isolation. In contrast, we focus on {\em system-level} verification, that is, verification of a closed-loop system consisting of a NN controller and an environment.

System-level verification approaches have also been proposed in the literature. A number of methods capture NNCSs as 
{\em hybrid systems}: in~\cite{ivanovVerisigVerifyingSafety2019}  the NN is transformed into a hybrid system and the existing tool Flow*~\cite{chenFlowAnalyzerNonlinear2013} is used to verify the resulting hybrid system. In~\cite{huangReachNNReachabilityAnalysis} the NN is approximated using Bernstein polynomials, while~\cite{verisig20} employs Taylor models, and uses various techniques to reduce error. 
In~\cite{eliyahu2021verifying} NNCSs are modeled as transition systems, and existing NN verifiers like Marabou~\cite{katz2019marabou} are used for bounded model checking.
These methods are adept at proving safety properties within a bounded time horizon. 
The ARCH-COMP series of competitions includes categories for verifying continuous and hybrid systems with neural network components~\cite{ARCH23:ARCH_COMP23_Category_Report_Artificial}. The tools~\cite{althoff2015_CORA_introduction, bogomolov2019juliareach, tran2020nnv, huang2022polar,ivanovVerisigVerifyingSafety2019,fanReachNNToolReachability2020,dutta2019sherlock} that participated in these competitions are designed to verify properties within a bounded time horizon by reachability analysis. 
In contrast, our approach is designed to verify safety properties over an infinite time horizon 
using inductive invariants instead of reachability.
Also, our approach works for a discrete-time rather than continuous-time model.
In~\cite{bacci2021verifying_rl_infinity}, the authors perform time-unbounded safety verification of NNCS within a conventional reinforcement learning setting that features a finite set of possible actions by computing polyhedral overapproximations of the reach set using MILP techniques. In contrast, our method does not assume a finite action space and uses inductive invariants.

Our approach can handle non-deterministic environments. While the non-deterministic case is less studied than deterministic NNCSs, some of the tools mentioned above, 
e.g.,~\cite{tran2020nnv,althoff2015_CORA_introduction}, can also handle non-determinism,
albeit with a fundamentally different technique than ours (reachability instead of inductive invariants).
The verification of non-deterministic NNCSs has also been studied in~\cite{akintunde2022non_det_neural_agents}, which leverages MILP solvers for the verification task.

In addition to verification, various methods for testing NNCSs have been explored.
Simulation-based approaches for NNCS analysis, including falsification, fuzz testing, and counterexample analysis, were introduced by~\cite{verifai-cav19,viswanadha2021parallelscenicverifai}.
\cite{goyalNeuralExplorer2020} proposed a method to explore the state space of hybrid systems containing neural networks.

Our work is also related to research on automatic inductive invariant discovery, which is a hard, generally undecidable, problem~\cite{padon2016decidabilityIndInv}. Recently, several studies have proposed techniques for automatic inductive invariant discovery for distributed protocols~\cite{goel2021symmetry,schultz2022plaininvariantinference}, while~\cite{ryan2019cln2inv} applies deep learning techniques to infer loop invariants for programs. 
\cite{amir2021towards} proposes a heuristic for inferring simple inductive invariants in NNCSs, within the context of communication networking systems. Such systems possess unique traits not commonly found in general NNCSs. These specific traits enable using a NN verifier to check a weaker condition instead of directly verifying the inductiveness condition.
In contrast, our method does not depend on specific properties of specialized NNCSs.

Our work is also related to {\em barrier certificates}, an alternative technique for verifying continuous-time systems~\cite{prajna2004barrier_certificates, sogokon2018vector_lyapunov}.
In~\cite{wang2021synthesizing_barrier_certificates} the authors synthesize barrier certificates by solving an optimization problem subject to bilinear matrix inequalities.
In~\cite{deshmukh2019learning_barrier} the authors propose a method that uses barrier certificates to synthesize neural network controllers which are safe by design. 
\cite{sha2021synthesizing_barrier_certificates_nncs} presents an approach to verifying NNCSs by constructing barrier certificates.

    \section{Conclusions}
    \label{sec:conclusion}
    We present a compositional, inductive-invariant based method for NNCS verification.
Our method allows to to verify safety properties over an infinite time horizon.
The key idea is to decompose the monolithic inductiveness check (which is typically not supported by state-of-the-art NN verifiers, and does not scale with state-of-the-art SMT solvers) into several manageable subproblems, which can be each individually handled by the corresponding tool. 
Our case studies show encouraging results where the verification time is reduced from hours (or timeout) to seconds.

Future work includes augmenting our method's capabilities to suit a broader range of NNCS applications. We also plan to explore the automatic generation of candidate inductive invariants (in addition to bridges) for NNCSs.

\section*{Acknowledgments}
We would like to thank the anonymous Reviewers of the submitted version of this paper for their helpful feedback. This work has been supported in part by NSF CCF award \#2319500.

    \bibliography{ref}

    \appendix
    \section*{Appendix}
    \section{Experiments with JuliaReach and NNV}
\label{sec:appendix_other_tools_experiments}

As mentioned in Section~\ref{sec:experiments_implementation},
following the recommendation of an anonymous Reviewer of the conference version of this paper~\cite{zhouNFM2024},
we performed some experiments with NNCS reachability analysis tools, specifically,
with JuliaReach~\cite{bogomolov2019juliareach} and NNV~\cite{tran2020nnv}. 
In this section, we present a summary of these experimental results. 
We utilize the deterministic 2D maze case study, as discussed in Section~\ref{sec:experiments_implementation}, to assess the performance of these tools. All necessary files and instructions to reproduce the results are available at \url{https://github.com/YUH-Z/comp-indinv-verification-nncs}.

\subsection{Model conversion}

Our tool is designed to work with PyTorch models for the controller, with the environment modeled within a Python file. However, JuliaReach and NNV require different formats for both the controller and the environment. While the environment can be manually implemented in Julia and Matlab, converting the NN controller into a format compatible with JuliaReach and NNV is necessary. 

Figure~\ref{fig:exp_import_explain} illustrates the conversion process. The conversion begins by converting the pretrained PyTorch model into the ONNX format, utilizing PyTorch's native ONNX exporter. NNV then employs its ONNX importer to translate the ONNX model into NNV's proprietary format. Additionally, NNV offers a discrete environment model, allowing us to represent our environment directly. With the model and environment thus prepared, we proceed to perform a 50-step reachability analysis.

JuliaReach also offers an ONNX importer capable of converting a limited subset of ONNX models into JuliaReach's internal format. Unfortunately, the ONNX models exported directly from our PyTorch models are incompatible with this importer. Consequently, we developed a converter to transform the ONNX files into NNet files, a format compatible with JuliaReach. While the NNet repository\footnote{\url{https://github.com/sisl/NNet}} does contain an ONNX converter, it does not support our ONNX models due to the presence of unsupported operators. Our converter builds upon NNet's existing ONNX converter, extending its functionality to support these operators. 

We evaluated the ONNX to NNet converter by comparing the output of the ONNX model with that of the NNet model converted from the same ONNX model. To conduct the assessment, we generated one million uniformly distributed input vectors for each NN controller configuration. Subsequently, we computed the absolute difference between the outputs of the ONNX and NNet models for each input vector. The largest difference observed across all NN controller configurations and input vectors was on the order of $10^{-6}$, which we considered acceptable for our purposes. 
(The origin of this difference is unclear.
Considering that the converter maintains consistent precision in floating-point representation, it is possible that the difference in the output values arises from differences in operator implementation between the ONNX and NNet libraries.)

To model the environment in JuliaReach, which does not natively support reachability analysis on discrete-time systems, we represented the deterministic 2D maze as a continuous-time system, aligning with the continuous-discrete conversion strategy used in ARCH-COMP~\cite{ARCH23:ARCH_COMP23_Category_Report_Artificial}. This continuous-time system has a time increment of $1.0$ and the following dynamics:
\begin{align}
\dot{x} = 0.1a, \quad \dot{y} = 0.1b,
\end{align}
where $x$ and $y$ represent the state variables, and $a$ and $b$ are control inputs produced by the NN controller. 
Following a procedure similar to the NNV setup, we conducted a reachability analysis on this system from time $0$ to time $50$.

\pgfdeclarelayer{background}
\pgfdeclarelayer{foreground}
\pgfsetlayers{background,main,foreground}

\tikzstyle{sensor}=[draw, fill=blue!20, text width=6em, 
    text centered, minimum height=2.5em,drop shadow]
\tikzstyle{ann} = [above, text width=5em, text centered]
\tikzstyle{wa} = [sensor, text width=10em, fill=red!20, 
    minimum height=6em, rounded corners, drop shadow]
\tikzstyle{sc} = [sensor, text width=13em, fill=red!20, 
    minimum height=10em, rounded corners, drop shadow]
\tikzstyle{util} = [sensor, text width=6em, fill=green!20, 
    minimum height=6em, rounded corners, drop shadow]

\def\blockdist{2.3}
\def\edgedist{2.5}

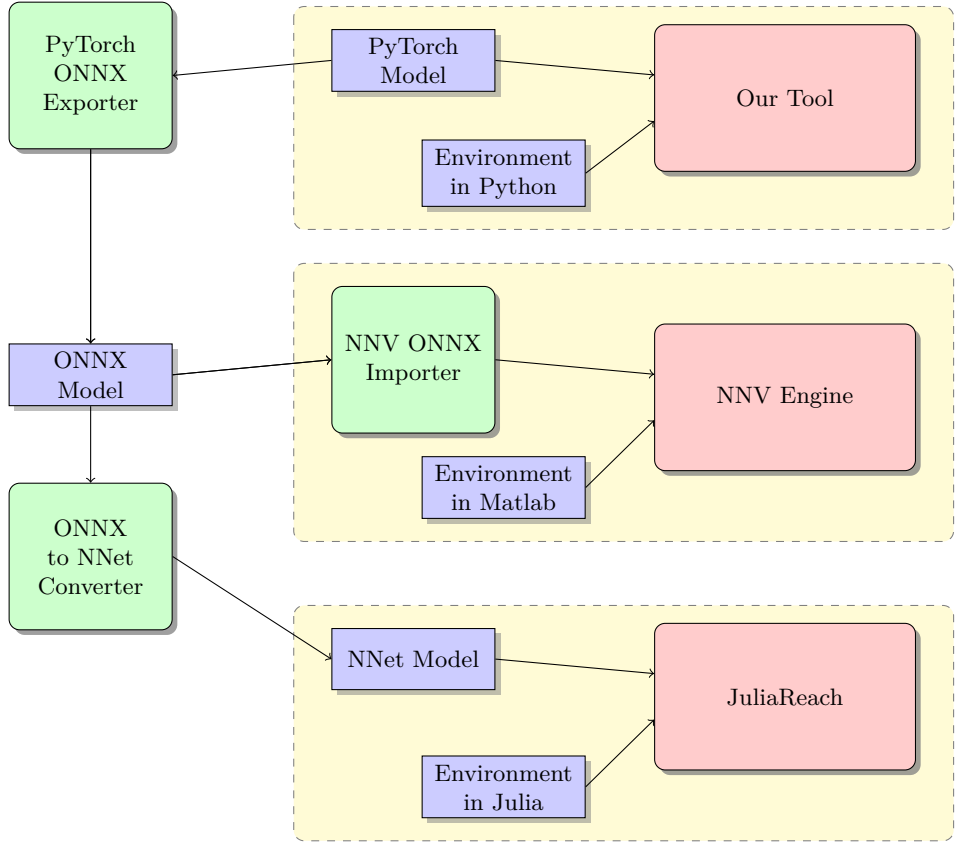
\begin{figure}[h]
  \centering
  \begin{tikzpicture}
    \node (wa) [wa]  {Our Tool};
    \path (wa.west)+(-3.2,0.5) node (asr1) [sensor] {PyTorch Model};
    \path (wa.west)+(-2.0,-1.0) node (asr2) [sensor] {Environment in Python};
    \path (asr1.west)+(-3.2,-0.2) node (onnxexport) [util] {PyTorch ONNX Exporter};

    \path [draw, ->] (asr1.east) -- node [above] {} 
        (wa.170) ;
    \path [draw, ->] (asr2.east) -- node [above] {} 
        (wa.190);
    \path [draw, ->] (asr1.west) -- node [above] {} (onnxexport.east);
  
    \begin{pgfonlayer}{background}
        \path (asr1.west |- asr1.north)+(-0.5,0.3) node (a) {};
        \path (asr2.south -| wa.east)+(+0.5,-0.3) node (b) {};
          
        \path[fill=yellow!20,rounded corners, draw=black!50, dashed]
            (a) rectangle (b);           
        \path (asr1.north west)+(-0.2,0.2) node (a) {};
    \end{pgfonlayer}
    
    \path (onnxexport.south)+(0.0, -3.0) node (onnx) [sensor] {ONNX Model};

    \path (wa.south)+(0.0, -3.0) node (wb) [wa] {NNV Engine};
    \path (wb.west)+(-3.2,0.5) node (bsr1) [util] {NNV ONNX Importer};
    \path (wb.west)+(-2.0,-1.2) node (bsr2) [sensor] {Environment in Matlab};

    \path [draw, ->] (bsr1.east) -- node [above] {} 
        (wb.170) ;
    \path [draw, ->] (bsr2.east) -- node [above] {} 
        (wb.190);
    \path [draw, ->] (onnxexport.south) -- node [above] {} (onnx.north);
    \path [draw, ->] (onnx.east) -- node [above] {} (bsr1.west);
  
    \begin{pgfonlayer}{background}
        \path (bsr1.west |- bsr1.north)+(-0.5,0.3) node (a) {};
        \path (bsr2.south -| wb.east)+(+0.5,-0.3) node (b) {};
          
        \path[fill=yellow!20,rounded corners, draw=black!50, dashed]
            (a) rectangle (b);           
        \path (bsr1.north west)+(-0.2,0.2) node (a) {};
    \end{pgfonlayer}

    \path (onnx.south)+(0.0, -2.0) node (onnx2nnet) [util] {ONNX to NNet Converter};

    \path (wb.south)+(0.0, -3.0) node (wc) [wa] {JuliaReach};
    \path (wc.west)+(-3.2,0.5) node (csr1) [sensor] {NNet Model};
    \path (wc.west)+(-2.0,-1.2) node (csr2) [sensor] {Environment in Julia};

    \path [draw, ->] (csr1.east) -- node [above] {} 
        (wc.170) ;
    \path [draw, ->] (csr2.east) -- node [above] {} 
        (wc.190);
    \path [draw, ->] (onnxexport.south) -- node [above] {} (onnx.north);
    \path [draw, ->] (onnx.east) -- node [above] {} (bsr1.west);
  
    \path [draw, ->] (onnx2nnet.east) -- node [above] {} (csr1.west);
    \path [draw, ->] (onnx.south) -- node [above] {} (onnx2nnet.north);
    \begin{pgfonlayer}{background}
        \path (csr1.west |- csr1.north)+(-0.5,0.3) node (a) {};
        \path (csr2.south -| wc.east)+(+0.5,-0.3) node (b) {};
          
        \path[fill=yellow!20,rounded corners, draw=black!50, dashed]
            (a) rectangle (b);           
        \path (csr1.north west)+(-0.2,0.2) node (a) {};
    \end{pgfonlayer}
  \end{tikzpicture}
  \caption{Overview of converting our models into formats compatible with NNV and JuliaReach.}
  \label{fig:exp_import_explain}
\end{figure}

\subsection{Experimental results}

The experimental results using NNV and JuliaReach are shown in Figures~\ref{fig:nnv_results} and~\ref{fig:juliareach_results}, respectively. In these figures, the green regions indicate the area deemed safe, the orange regions indicate the initial states, and the yellow regions represent the (approximated) reach sets computed by the tools. In every scenario, both NNV and JuliaReach completed the reachability analysis within a few seconds. 

NNV failed to verify safety for NN controllers with configurations of $2\times 40$ and $2\times 512$ neurons. Specifically, as shown in Figures~\ref{fig:nnv_40_40} and~\ref{fig:nnv_512_512}, the reach sets computed by NNV are not subsets of the safe regions. This is evident in the top right corners of the figures, where the yellow boxes exceed the boundaries of the green region. Consequently, NNV cannot prove that all reachable states, within a maximum of 50 steps from the initial state, satisfy the safety property, even though this is indeed the case (as the inductive invariant method shows).

JuliaReach failed to verify safety for NN controllers with configurations of $2\times 512$ and $2\times 1024$ neurons. In both Figures~\ref{fig:juliareach_512_512} and~\ref{fig:juliareach_1024_1024}, the yellow boxes exceed the  boundaries of the green regions. 

The inability of NNV and JuliaReach to prove safety in some cases is attributed to the over-approximation errors introduced by the reachability analysis performed by these tools. 
In contrast, the inductive invariant method is exact. As detailed in Section~\ref{sec:experiments_implementation}, our tool managed to verify safety across all tested scenarios in a few seconds, demonstrating the effectiveness of our proposed method.

\begin{figure}[ht]
  \centering
  \subfloat[$2 \times 32$]{\includegraphics[width=0.75\textwidth]{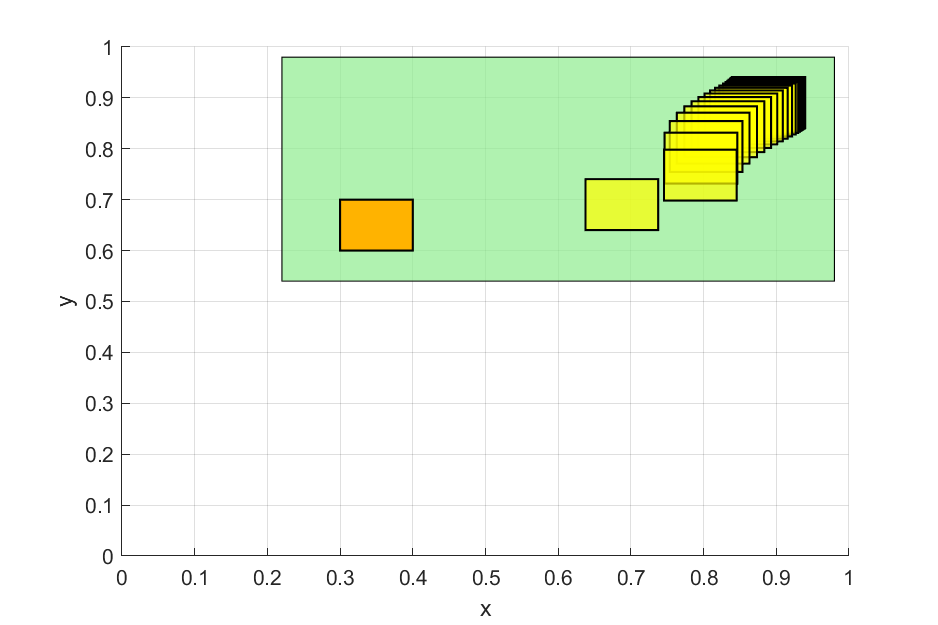}\label{fig:nnv_32_32}}

  \subfloat[$2 \times 40$]{\includegraphics[width=0.75\textwidth]{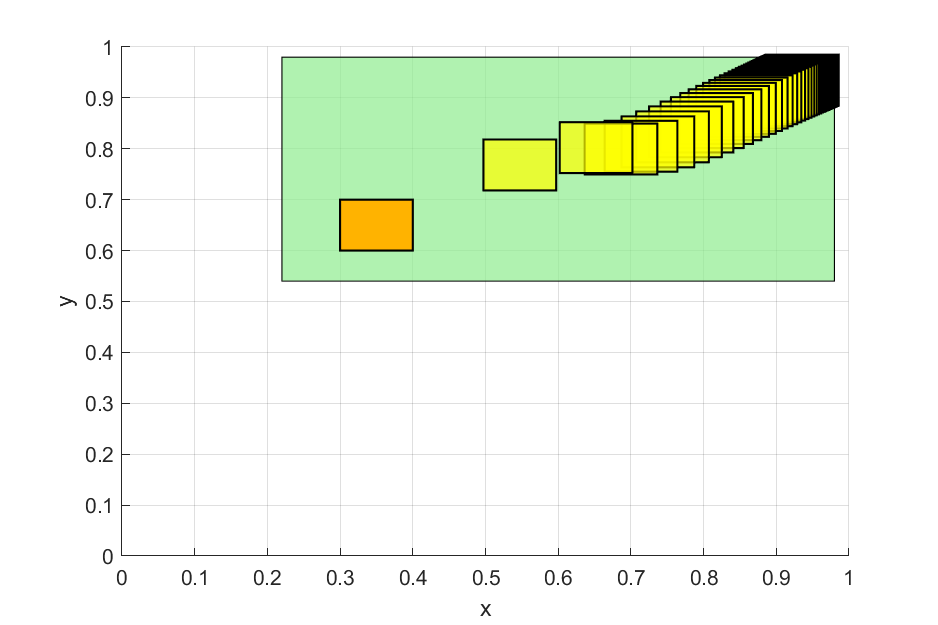}\label{fig:nnv_40_40}}

  \subfloat[$2 \times 48$]{\includegraphics[width=0.75\textwidth]{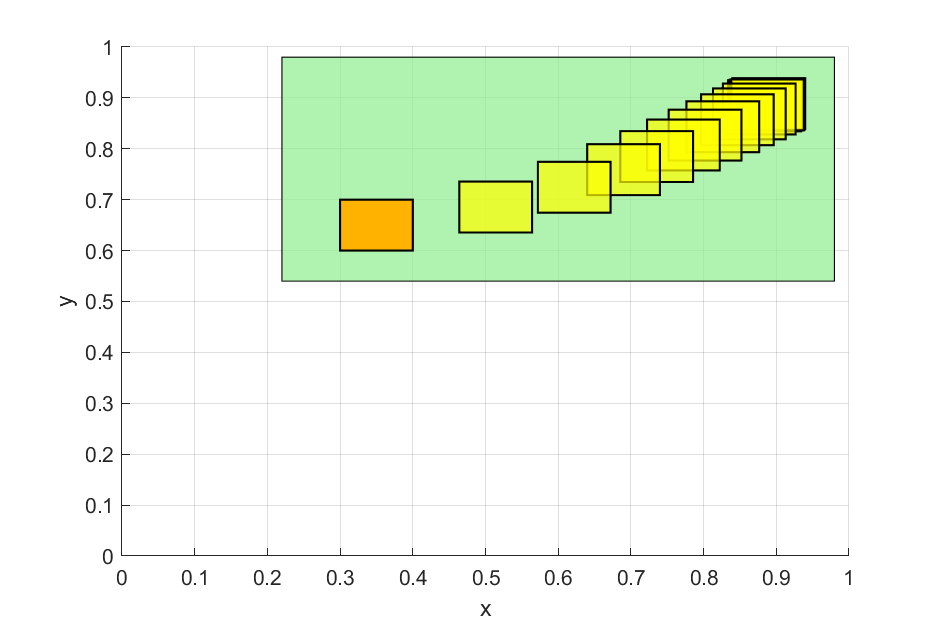}\label{fig:nnv_48_48}}

  \caption{NNV reachability analysis results for deterministic 2D maze, with subcaptions denoting the number of neurons in the NN controllers.}
  \label{fig:nnv_results}
\end{figure}

\begin{figure}[ht]\ContinuedFloat
  \centering
  \subfloat[$2 \times 56$]{\includegraphics[width=0.75\textwidth]{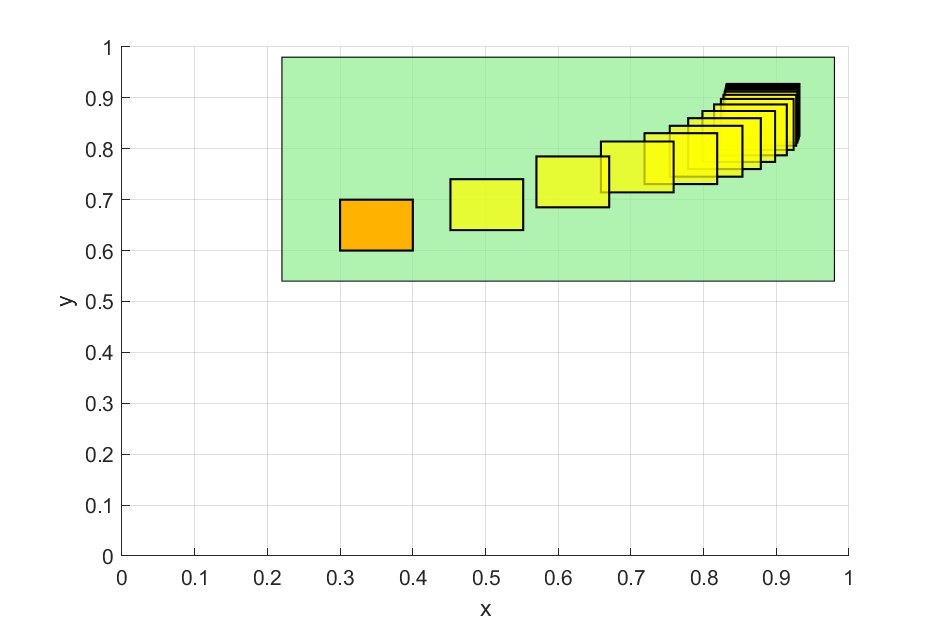}\label{fig:nnv_56_56}}

  \subfloat[$2 \times 64$]{\includegraphics[width=0.75\textwidth]{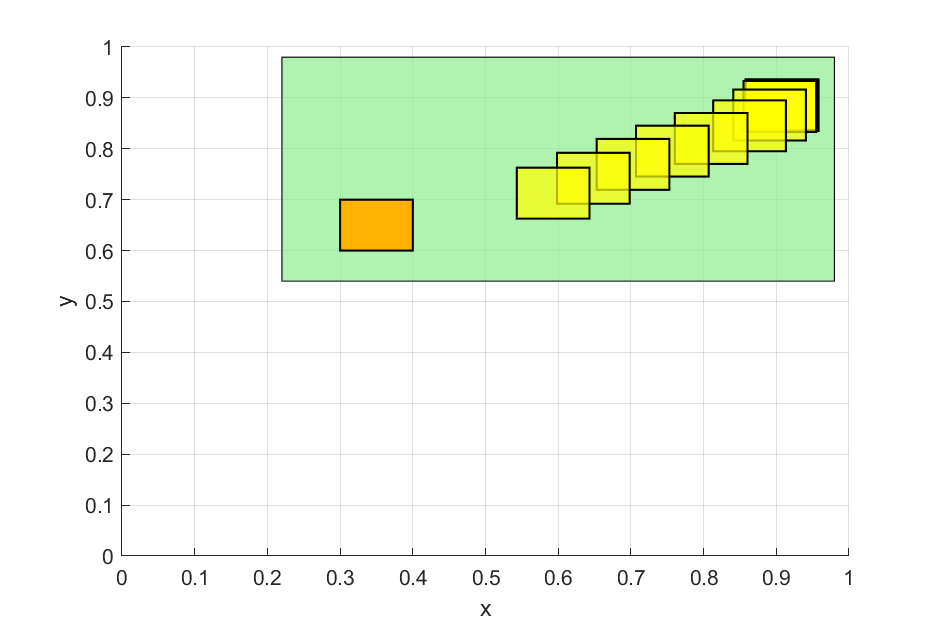}\label{fig:nnv_64_64}}
  
  \subfloat[$2 \times 128$]{\includegraphics[width=0.75\textwidth]{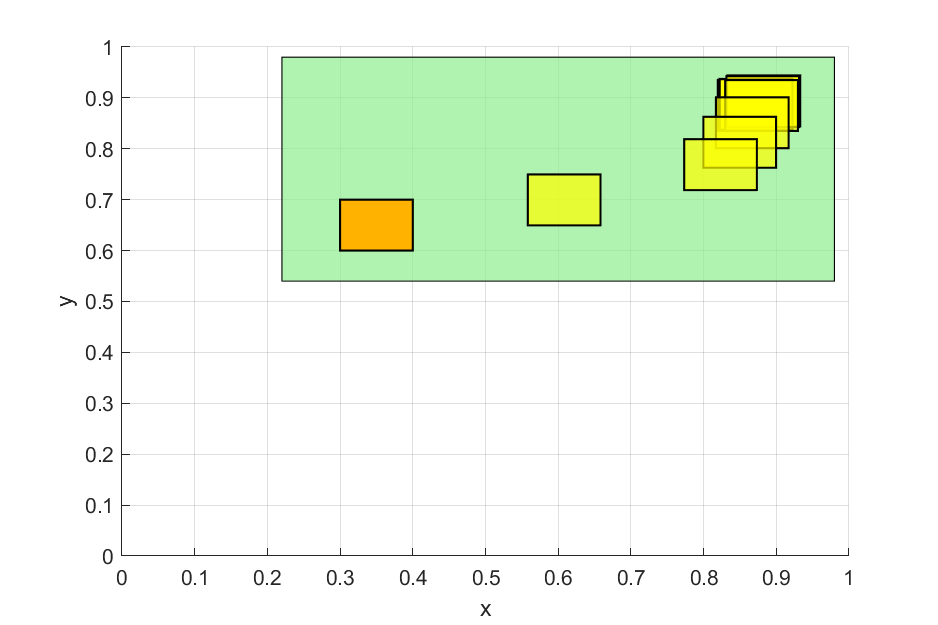}\label{fig:nnv_128_128}}
  \caption{NNV reachability analysis results for deterministic 2D maze, with subcaptions denoting the number of neurons in the NN controllers.}
\end{figure}

\begin{figure}[ht]\ContinuedFloat
  \centering
  \subfloat[$2 \times 256$]{\includegraphics[width=0.75\textwidth]{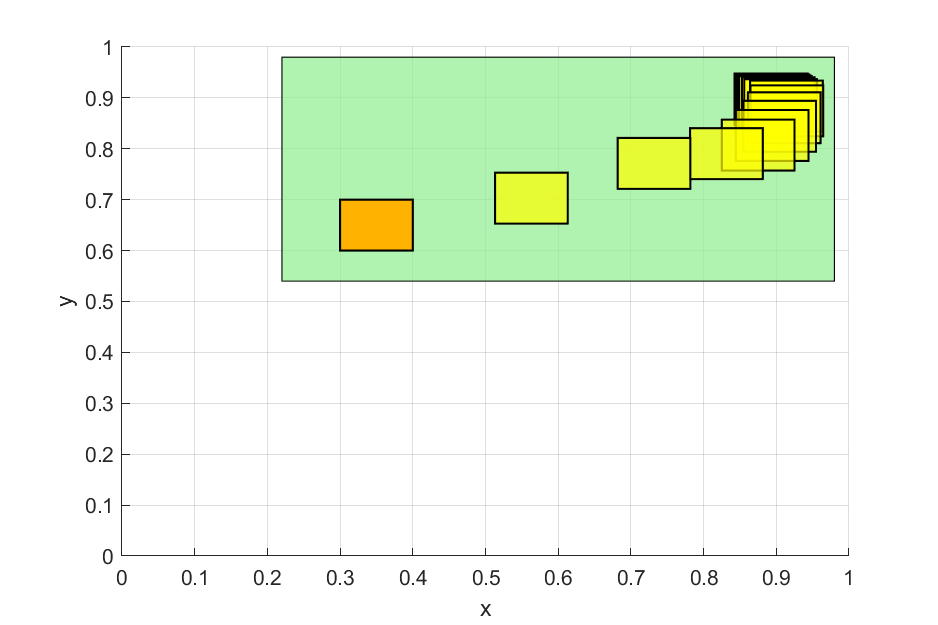}\label{fig:nnv_256_256}}
  
  \subfloat[$2 \times 512$]{\includegraphics[width=0.75\textwidth]{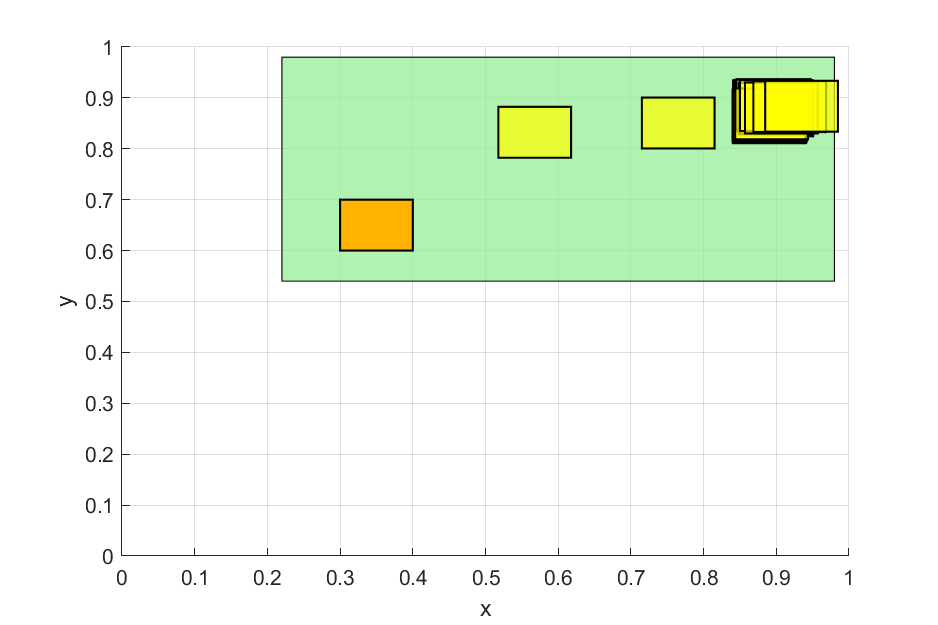}\label{fig:nnv_512_512}}

  \subfloat[$2 \times 1024$]{\includegraphics[width=0.75\textwidth]{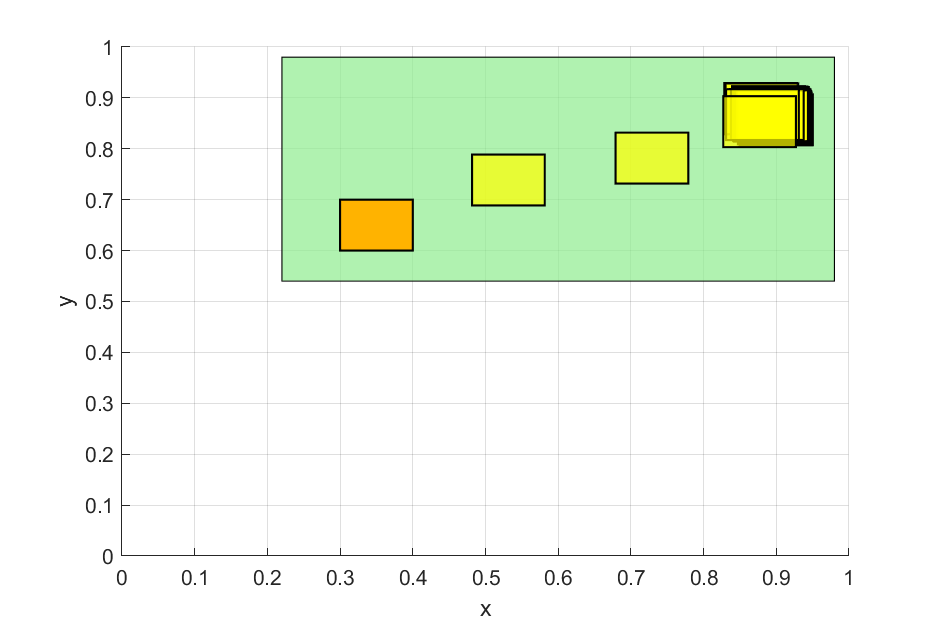}\label{fig:nnv_1024_1024}}
  \caption{NNV reachability analysis results for deterministic 2D maze, with subcaptions denoting the number of neurons in the NN controllers.}
\end{figure}

\begin{figure}[ht]
  \centering
  \subfloat[$2 \times 32$]{\includegraphics[width=0.75\textwidth]{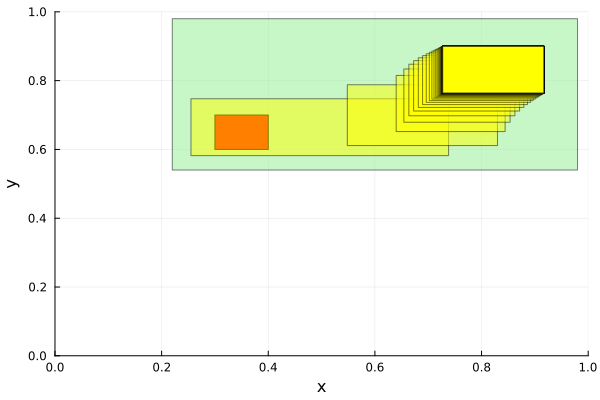}\label{fig:juliareach_32_32}}

  \subfloat[$2 \times 40$]{\includegraphics[width=0.75\textwidth]{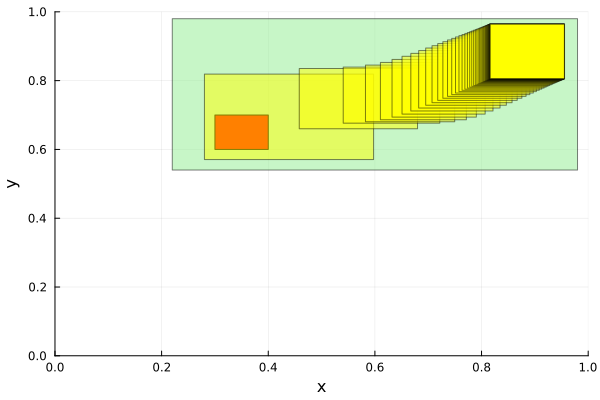}\label{fig:juliareach_40_40}}

  \subfloat[$2 \times 48$]{\includegraphics[width=0.75\textwidth]{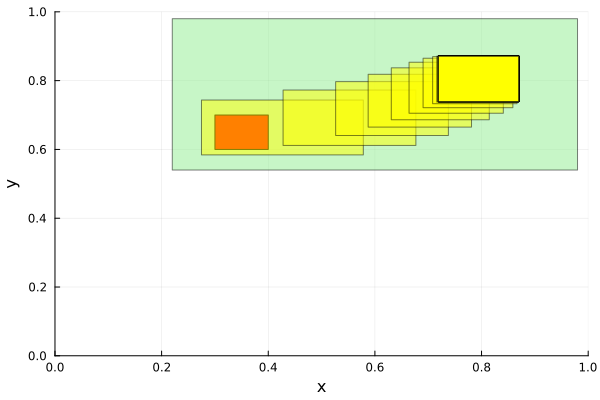}\label{fig:juliareach_48_48}}

  \caption{JuliaReach reachability analysis results for deterministic 2D maze, with subcaptions denoting the number of neurons in the NN controllers.}
  \label{fig:juliareach_results}
\end{figure}

\begin{figure}[ht]\ContinuedFloat
  \centering
  \subfloat[$2 \times 56$]{\includegraphics[width=0.75\textwidth]{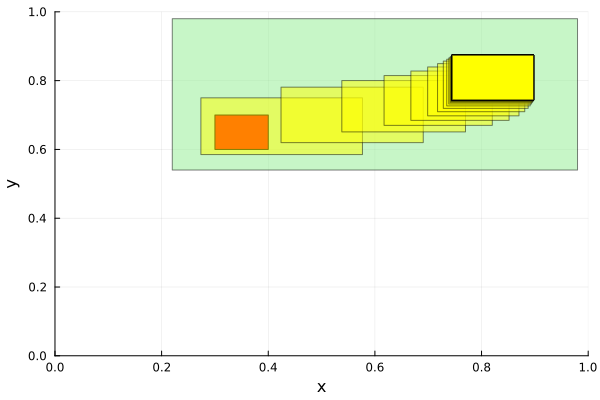}\label{fig:juliareach_56_56}}

  \subfloat[$2 \times 64$]{\includegraphics[width=0.75\textwidth]{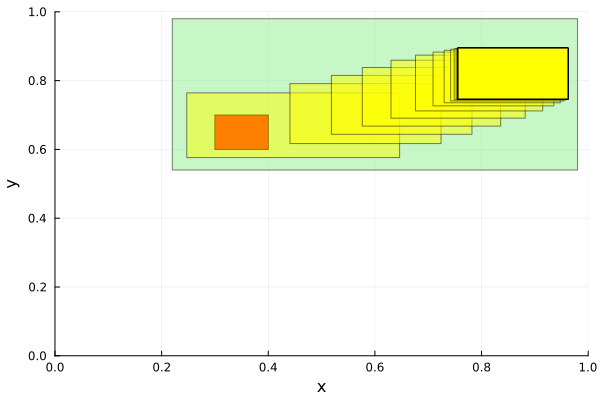}\label{fig:juliareach_64_64}}
  
  \subfloat[$2 \times 128$]{\includegraphics[width=0.75\textwidth]{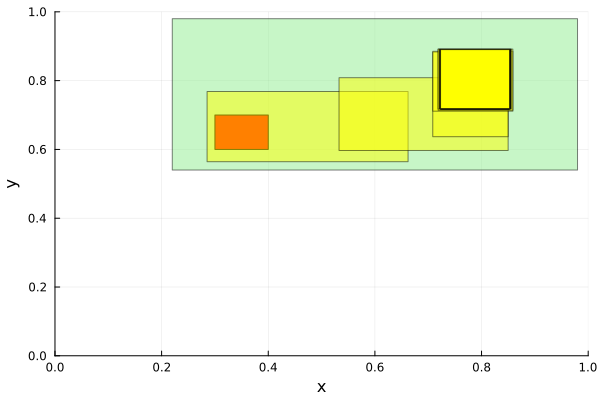}\label{fig:juliareach_128_128}}
  \caption{JuliaReach reachability analysis results for deterministic 2D maze, with subcaptions denoting the number of neurons in the NN controllers.}
\end{figure}

\begin{figure}[ht]\ContinuedFloat
  \centering
  \subfloat[$2 \times 256$]{\includegraphics[width=0.75\textwidth]{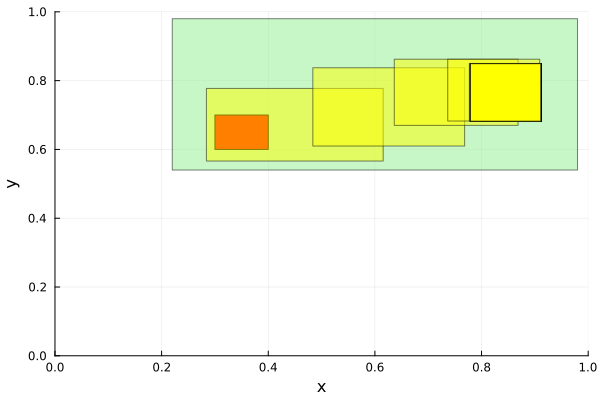}\label{fig:juliareach_256_256}}
  
  \subfloat[$2 \times 512$]{\includegraphics[width=0.75\textwidth]{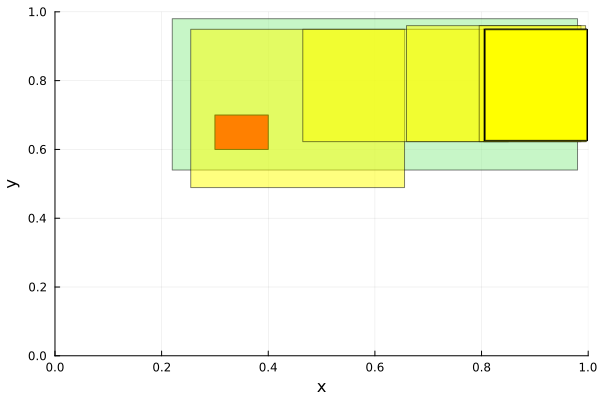}\label{fig:juliareach_512_512}}

  \subfloat[$2 \times 1024$]{\includegraphics[width=0.75\textwidth]{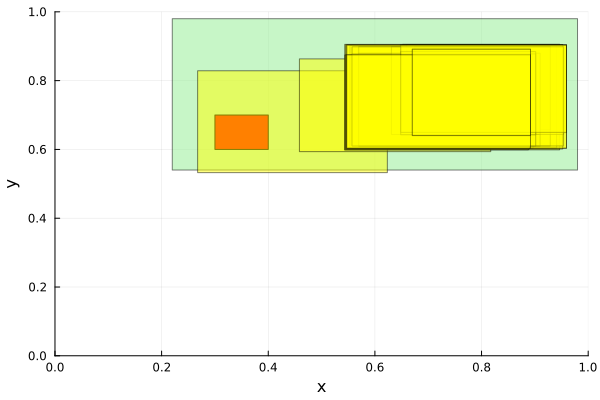}\label{fig:juliareach_1024_1024}}
  \caption{JuliaReach reachability analysis results for deterministic 2D maze, with subcaptions denoting the number of neurons in the NN controllers.}
\end{figure}

\end{document}